\documentclass[iop]{emulateapj}

\usepackage{url}

\newcommand{\rmxaa}{RevMexAA}

\shortauthors{Mackey et al.}
\shorttitle{Double bow shocks around young, runaway red supergiants}

\begin{document}

\title{Double bow shocks around young, runaway red supergiants: application to Betelgeuse}

\author{Jonathan Mackey\altaffilmark{1}}
\altaffiltext{1}{Argelander-Institut f\"ur Astronomie, Auf dem H\"ugel 71, 53121 Bonn, Germany}
\email{jmackey@astro.uni-bonn.de}
\author{Shazrene Mohamed\altaffilmark{1,2}}
\altaffiltext{2}{South African Astronomical Observatory, P.O.\ box 9, 7935 Observatory, South Africa}
\author{Hilding R.\ Neilson\altaffilmark{1}}
\author{Norbert Langer\altaffilmark{1}}
\author{Dominique M.-A.\ Meyer\altaffilmark{1}}

\begin{abstract}
A significant fraction of massive stars are moving supersonically through the interstellar medium (ISM), either due to disruption of a binary system or ejection from their parent star cluster.  The interaction of their wind with the ISM produces a bow shock.  In late evolutionary stages these stars may undergo rapid transitions from red to blue and vice versa on the Hertzsprung-Russell diagram, with accompanying rapid changes to their stellar winds and bow shocks.  Recent 3D simulations of the bow shock produced by the nearby runaway red supergiant (RSG) Betelgeuse, under the assumption of a constant wind, indicate that the bow shock is very young ($<30\,000$ years old), hence Betelgeuse may have only recently become a RSG.  To test this possibility, we have calculated stellar evolution models for single stars which match the observed properties of Betelgeuse in the RSG phase.  The resulting evolving stellar wind is incorporated into 2D hydrodynamic simulations in which we model a runaway blue supergiant (BSG) as it undergoes the transition to a RSG near the end of its life.  We find that the collapsing BSG wind bubble induces a bow shock-shaped inner shell around the RSG wind that resembles Betelgeuse's bow shock, and has a similar mass.  Surrounding this is the larger-scale retreating bow shock generated by the now defunct BSG wind's interaction with the ISM.  We suggest that this outer shell could explain the bar feature located (at least in projection) just in front of Betelgeuse's bow shock.
\end{abstract}


\section{Introduction}
\label{sec:intro}
Winds from massive stars have a strong effect on their circumstellar environment by driving shocks and producing parsec-scale shells~\citep{DysdeV72,CasMcCWea75}.
When the star is moving these shells become asymmetric and, if the star is moving supersonically with respect to the interstellar medium (ISM), a bow shock is formed whose shape was calculated by \citet{Wil96}.
A significant fraction of massive O-stars become runaway stars because of dynamical ejection from clusters and binary disruption \citep[e.g.][]{GieBol86, PflKro06, EldLanTou11}.
Bow shocks around massive stars have been found in optical images~\citep{GulSof79} and with the \textit{IRAS} satellite in the mid- and far-infrared (IR) \citep{vBurMcC88,vBurNorCreDga95}.
IR observations are very efficient for detecting and characterising bow shocks~\citep[e.g.][]{GvaBom08, PovBenWhiEA08, WinWolBouEA12}.

\begin{figure}
\centering
\resizebox{\hsize}{!}{\includegraphics{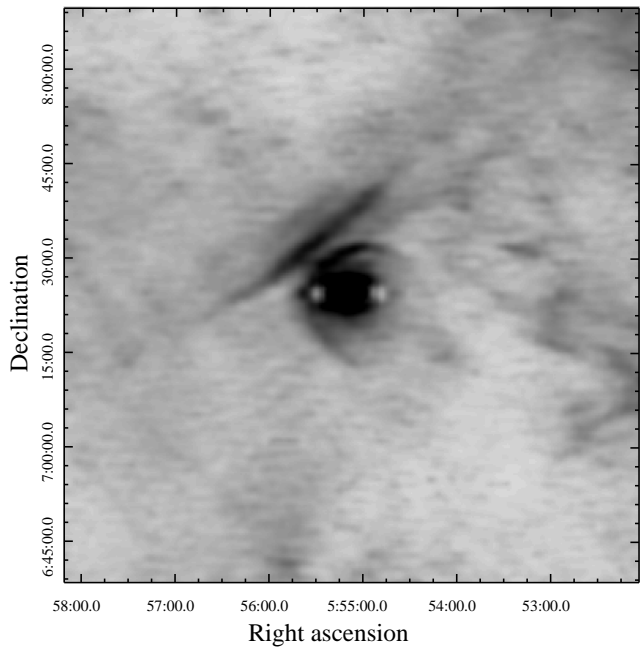}}
\caption{Betelgeuse's bow shock and ``bar'' at 60$\mu$m, from the \textit{IRAS} Galaxy Atlas \protect\citep{CaoTerPri97}.
  }
\label{fig:Betelgeuse}
\end{figure}

 \textit{IRAS} observations \citep{NorBurCaoEA97} of the red supergiant (RSG) Betelgeuse ($\alpha$ Orionis) detected a bow shock and a linear ``bar'' feature upstream from the bow shock and perpendicular to the star's proper motion, shown here in Fig.~\ref{fig:Betelgeuse}.
Higher resolution  \textit{AKARI}  observations \citep{UetIzuYamEA08} confirmed these features.  Recent \textit{Herschel} observations \citep{CoxKervMarEA12} showed layered substructure in the bow shock.
\citet{LeBMatGerEA12} found a thin shell of far-ultraviolet emission at the same position as the IR bow shock.
They also presented VLA 21cm observations with evidence of some atomic hydrogen near the bow shock, and stronger emission from an inner shell three times closer to Betelgeuse.

Thin shells and bow shocks are often subject to instabilities for realistic stellar winds and ISM parameters, requiring multi-dimensional simulations for quantitative studies \citep[e.g.][]{GarSegMacLowLan96}.
Runaway stars with constant winds have been extensively modelled for both fast winds from hot stars \citep{BriDErc95a, RagNorCreCanEA97, ComKap98} and slow winds from cool stars \citep[e.g.][and references therein]{VilManGar12,CoxKervMarEA12}.
Bow shocks produced by stars evolving from red to blue have been studied previously \citep{BriDErc95b, vMarLanAchEA06}, but there have been no detailed studies of the opposite transition from blue to red.

\citet{MohMacLan12} modelled Betelgeuse's bow shock using 3D SPH simulations assuming a constant stellar wind.  By comparing the observed mass ($M_s\simeq0.0033\,\mathrm{M}_{\sun}$ from \textit{AKARI} data) and shape of the bow shock with simulations they found that the bow shock must be very young ($\lesssim30$ kyr) and not yet in a steady state.  This implies that either Betelgeuse has recently moved into a region of the ISM with properties very different to where it came from, or that the wind itself has changed significantly.  If Betelgeuse had recently undergone a transition from a blue supergiant (BSG) or main sequence (MS) star to a RSG, then the bar feature could be a remnant from the MS or BSG wind \citep{MohMacLan12}.  In this paper we test this idea and investigate such a transition with 2D hydrodynamical simulations of an evolving wind.

\begin{figure}[t]
\centering
\resizebox{\hsize}{!}{\includegraphics{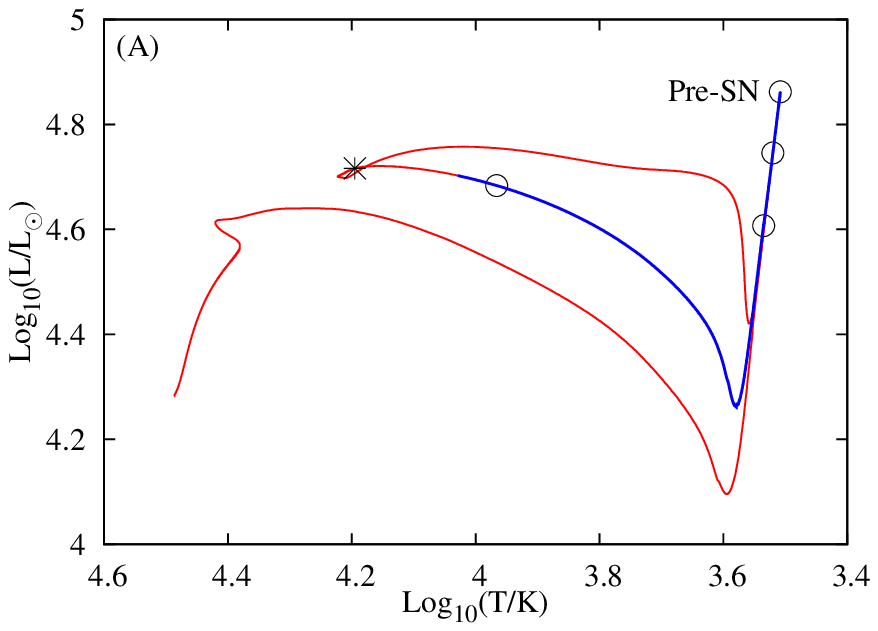}}
\resizebox{\hsize}{!}{\includegraphics{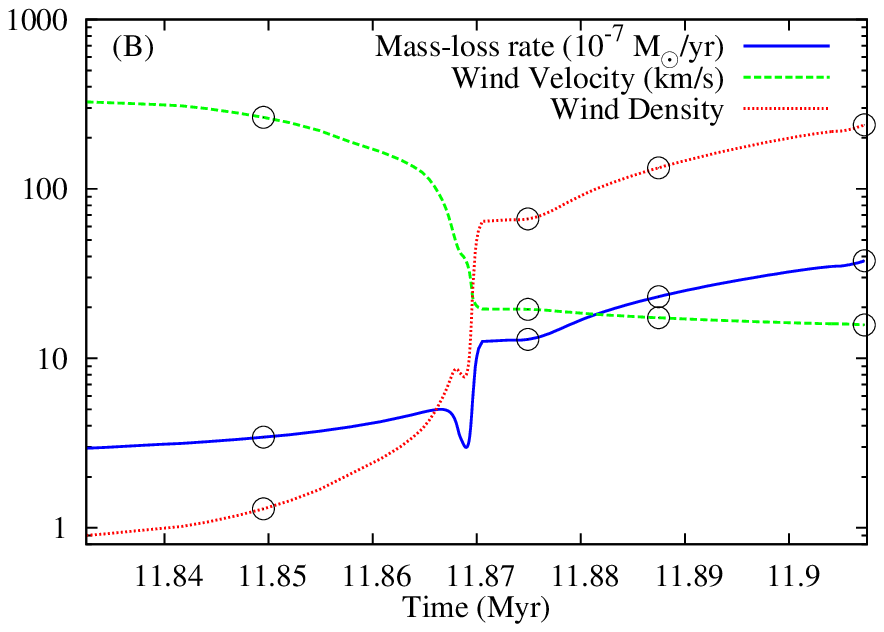}}
\caption{\textbf{(A)} Hertzsprung-Russell diagram for the $15\,\mathrm{M}_{\sun}$ stellar model.  The star symbol identifies $t=11.4$ Myr, midway through the extended blue loop and the starting point of the simulations in this paper.  The blue part of the curve highlights the final 75 kyr of evolution, shown in more detail in panel (B).  Open circles identify the times of the four snapshots in Fig.~\ref{fig:DensTemp}.  \textbf{(B)} The wind mass-loss rate, $\dot{M}$ in $10^{-7}\,\mathrm{M}_{\sun}\,\mathrm{yr}^{-1}$, wind velocity, $v_w$ in $\mathrm{km}\,\mathrm{s}^{-1}$, and  $\dot{M}/v_w$ (proportional to wind density) are shown.
  }
\label{fig:HRD}
\end{figure}

\section{Description of code and methods}
\label{sec:methods}

\subsection{Stellar evolution models}
We consider a 15 M$_\odot$ solar metallicity star whose evolution on the Hertzsprung-Russell diagram is shown in Fig.~\ref{fig:HRD}(A). The tip of the RSG branch has properties consistent with Betelgeuse \citep[][and references therein]{NeiLesHau11, SmiHinRyd09}.  The model is computed using the \citet{YooLan05} stellar evolution code including mass loss but no convective core overshooting.  Mass loss is computed using the \citet{KudPauPulEA89} approximation for hot stars and the \citet{JagNieHuc88} relation at cooler effective temperatures.
Wind velocities are approximated using the \citet{EldGenDaiEA06} prescription, $v_w^2 =  \beta_w(T) v_{\mathrm{esc}}^2$.
The value of of $\beta_w$ for $T<3600$ K was reduced from their value of $\beta_w=0.125$ to $\beta_w=0.04$ to match the observed wind velocity of Betelgeuse ($v_w\simeq17\,\mathrm{km}\,\mathrm{s}^{-1}$).
The wind velocity,  mass-loss rate $\dot{M}$, and wind density ($\dot{M}/v_w$), are plotted in Fig.~\ref{fig:HRD}(B) for the last $\sim75$ kyr of evolution. 
This corresponds to the end of the blue loop and evolution along the RSG branch, highlighted in blue in Fig.~\ref{fig:HRD}(A).

\subsection{Simulation setup}
The hydrodynamics code we have used \citep{MacLim10} is a finite-volume, uniform-grid code; here the equations of inviscid, compressible hydrodynamics are solved on an axisymmetric 2D grid in $(z,R)$, using a second-order-accurate integration scheme \citep{Fal91}.
Microphysical cooling processes provide a source term to the energy equation and are included in the code by operator splitting.
Cooling rates are calculated using the collisional ionisation equilibrium (CIE) cooling tables of \citet{WieSchSmi09} for solar abundances.
For numerical stability the gas temperature is limited to $T>30$ K.

The stellar wind boundary condition is implemented as in \citet{FreHenYor03} and \citet{vMarLanAchEA06} by imposing a freely expanding wind within a 20 grid-zone radius of the origin.
Wind parameters are updated from the stellar evolution model every timestep.
The initial ISM density and pressure were set to $\rho_0=4.676\times10^{-25}\,\mathrm{g}\,\mathrm{cm}^{-3}$ and $p_0=4.676\times10^{-13}\,\mathrm{dyne}\,\mathrm{cm}^{-2}$, respectively,  corresponding to a H number-density of $n_{\mathrm{H}}=0.2\,\mathrm{cm}^{-3}$ for $n_{\mathrm{H}}=\rho /2.338\times10^{-24} \,\mathrm{g}$ and a temperature $T=7700$ K for ionised gas containing 10\% helium by number.
The star's velocity through the ISM was set to $v_*=50\,\mathrm{km}\,\mathrm{s}^{-1}$.

The BSG phase has an increasing mass-loss rate of $\dot{M}\simeq(2\rightarrow3)\times10^{-7}\,\mathrm{M}_{\sun}\,\mathrm{yr}^{-1}$ as it approaches the transition to RSG, and a decreasing terminal velocity of $v_w\simeq400\rightarrow300\,\mathrm{km}\,\mathrm{s}^{-1}$ (see Fig.~\ref{fig:HRD}).  Equating the wind and ISM ram pressures gives the well-known standoff distance \citep[e.g.][]{Wil96}; $R_{\mathrm{SO}}\simeq0.6-0.7$ pc for the BSG wind just quoted.
The simulation domain was set to $z\in[-2.56,1.28]$ pc and $R\in[0,1.92]$ pc, using a coordinate system where the star is at the origin and the upstream direction is $+\hat{z}$; the code works in the reference frame where the star is stationary.
The upstream boundary condition was inflow, the symmetry axis was reflective, and the other boundary conditions were outflow-only.
The grid contained $1536\times768$ zones, giving a zone size $\delta x=0.0025$ pc and a wind-boundary radius of 0.05 pc.
A lower resolution simulation verified the numerical convergence of our results.
The simulation was started at $t=11.4$ Myr in the evolution model, about 400 kyr before the transition from BSG to RSG begins (see Fig.~\ref{fig:HRD}).
The ISM advects across the domain in 75 kyr, so 400 kyr is more than sufficient for any effects of the initial conditions to be erased.

\begin{figure*}
\centering
\resizebox{\hsize}{!}{\includegraphics{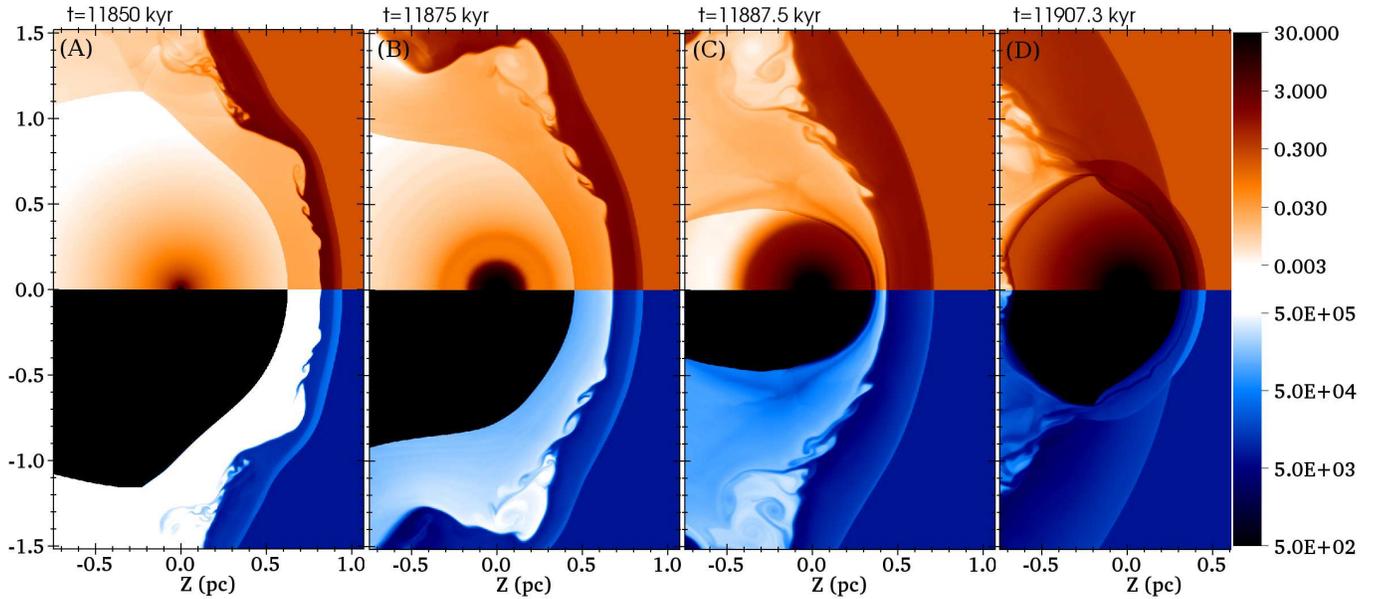}}
\caption{Gas number density $n_{\mathrm{H}}$ (upper half-plane, in $\mathrm{cm}^{-3}$) and temperature (lower half-plane, in Kelvin) plotted on logarithmic scales at times $t=11.85,\ 11.875,\ 11.8875,$ and $11.9073$ Myr from left to right, respectively.
Only part of the simulation domain is shown.}
\label{fig:DensTemp}
\end{figure*}

\begin{figure}
\centering
\resizebox{\hsize}{!}{\includegraphics{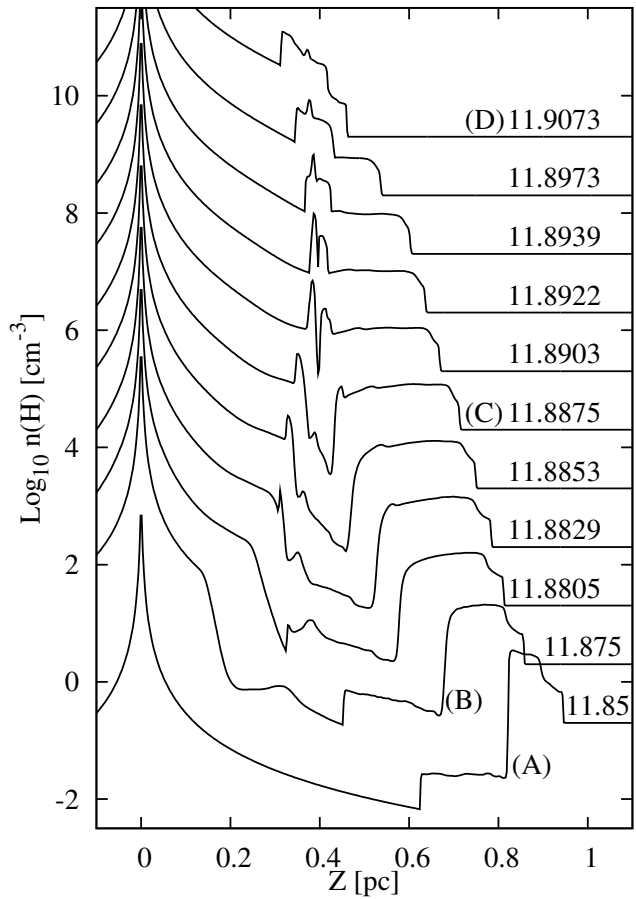}}
\caption{
  Gas density ($\mathrm{cm}^{-3}$) as a function of position (in pc, relative to the star) along the symmetry axis.
  Curves are labelled by the simulation time (in Myr) of the snapshot; each curve is offset by +1 dex from the previous one.
  Labels A-D correspond to the snapshots shown in panels A-D of Fig.~\ref{fig:DensTemp}.
  }
\label{fig:DensityCuts}
\end{figure}

\section{Results}
\label{sec:results}

Snapshots from the simulation showing $n_{\mathrm{H}}$ (above) and $T$ (below) are presented in Fig.~\ref{fig:DensTemp} at the evolutionary times indicated.
A more quantitative plot of the evolving density field along the symmetry axis is shown in Fig.~\ref{fig:DensityCuts}, where the four curves labelled A-D correspond to panels A-D of Fig.~\ref{fig:DensTemp}.
The first snaphot (A) shows the CSM at the end of the BSG evolutionary phase, with a single bow shock comprising the forward shock at $r=0.95$ pc, cooling-regulated compression in the shocked ISM at $r\simeq0.9$ pc, the contact discontinuity separating wind material from the ISM at $r=0.82$ pc, and the reverse/termination shock in the BSG wind at $r=0.63$ pc.
The forward shock is stable, with a density jump of $\simeq4\times$ across the shock, and a subsequent increase in density as post-shock gas cools.
Kelvin-Helmholtz instabilities are excited by shear across the contact discontinuity.
Further from the symmetry axis these instabilities can significantly affect the bow shock position and curvature.

During the BSG phase the shocked wind is approximately adiabatic, but the shocked ISM lies between the adiabatic and isothermal limits for the cooling function and ISM density considered.
The strong contact discontinuity ensures that the shell thickness is set only by the cooling of the shocked ISM layer.
Without cooling the shell is $\approx3\times$ thicker; with enhanced cooling it is thinner and very unstable.
Our model appears to use similar cooling to \citet{WarZijOBr07} but possibly somewhat weaker than \citet{VilManGar12}.
The uncertain balance between heating and cooling at $T\lesssim10^{4}\,$K at these low densities makes the shell thickness difficult to model reliably; additionally magnetic pressure could also play a role.

At time (B), the wind density has increased at least tenfold out to $r\sim0.15$ pc and there is a shell with enhanced density in the wind at $r\simeq0.3$ pc from the earlier local maximum in the wind density at $t=11.868$ Myr (Fig.~\ref{fig:HRD}B).
The BSG bow shock is no longer supported by the fast wind's ram pressure, so it retreats towards the slowly expanding RSG wind, preceded by the reverse shock.
The RSG wind has no shell because $v_w$ decreases monotonically with time during the transition (so successive layers of wind material don't interact).

Between times (B) and (C) the hot bubble still has a large sound speed, so the reverse shock collapses onto the RSG wind rapidly, compressing the shell at $r \simeq 0.3~$pc, and creating a thinner, denser shell at the edge of the RSG wind by 11.8829 Myr.
The mass of the inner shocked-shell grows linearly with time from zero at 11.882 Myr to $0.002\,\mathrm{M}_{\sun}$ at 11.892 Myr, when it reaches the upstream contact discontinuity and merges with it.

At time (C) the thin shell surrounding the RSG wind is shaped like a bow shock, with mass and physical size similar to that determined from \textit{AKARI} data \citep{MohMacLan12}. 
A second, much weaker shell is also seen at a slightly larger radius.
Both the BSG forward shock and contact discontinuity are still being swept downstream, but the weak shell-like feature at $z=0.44$ pc in curve C shows that gas pressure in the compressed hot bubble is impacting the contact discontinuity.
We identify the retreating contact discontinuity as a counterpart to the ``bar'' upstream from Betelgeuse's bow shock.
This transitory phase, from when the RSG wind encounters the BSG reverse shock to when it reaches the BSG contact discontinuity, lasts 10 kyr.
The two shells have merged by $t=11.8939$ Myr.

After time (C) the density structure is that of layered shells, which soon merge as the RSG wind pushes through the BSG bow shock, and by the end of the star's life at time (D) the RSG wind has reached the undisturbed ISM and is forming a new bow shock.
The RSG wind is also bounded by a shell downstream because of the pressure of the collapsing BSG bubble.
It would take $\sim60\,$kyr (from the time it loses pressure support) for the BSG shock to be completely swept downstream on the simulation domain shown, so the transition phase is far from complete at the pre-SN stage.

\section{Discussion and comparison to Betelgeuse} \label{sec:discussion}

\begin{figure}[t]
\centering
\resizebox{\hsize}{!}{\includegraphics{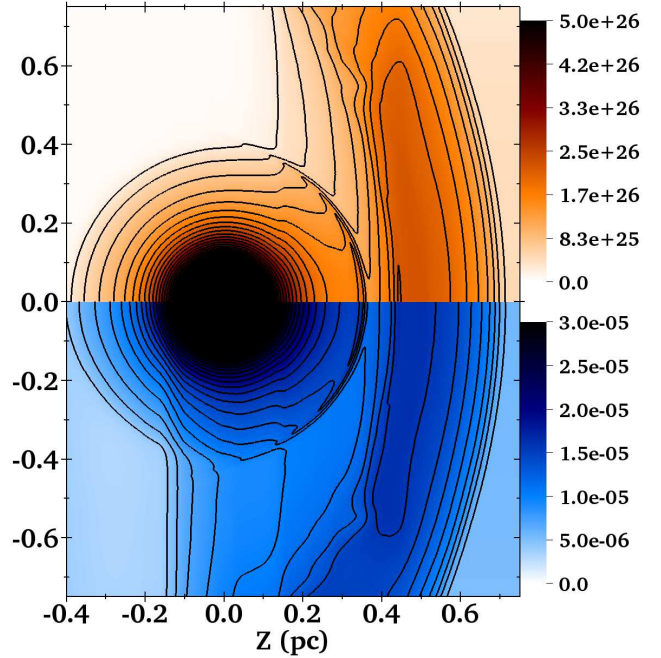}}
\caption{Projected dust luminosity  in the IRAS 65 $\mu$m band (above) and projected mass density (below) at $t=11.8875$ Myr.  The dust luminosity is shown in units of $\mathrm{erg}\,\mathrm{s}^{-1}$ per pixel, and the projected mass in $\mathrm{g}\,\mathrm{cm}^{-2}$, both on a linear scale.  20 contours are overlaid using the same linear scale.  Not all of the simulation domain is shown.}
\label{fig:Projected}
\end{figure}

IR observations of Betelgeuse's CSM \citep{NorBurCaoEA97, UetIzuYamEA08, CoxKervMarEA12} show that the standoff distance of the bow shock is  $\approx 0.3$ pc, and that of the bar is $\approx$ 0.45 pc (for $d = 200$ pc).  Fig.~\ref{fig:DensTemp} shows that a BSG bow shock which collapses as the star becomes a RSG produces strikingly similar structures.  In this model the ``bow shock'' around Betelgeuse is actually the collapsing BSG reverse shock interacting with the expanding RSG wind, and the bar is the BSG contact discontinuity.

The total gas mass in Betelgeuse's combined wind and bow shock is estimated at $0.033\,\mathrm{M}_{\sun}$ from 
\textit{IRAS} data \citep{MohMacLan12}; in our model at time (C) the total mass within $r<0.3$ pc is $0.030\,\mathrm{M}_{\sun}$, in good agreement.
The bow shock mass is $M_s\sim0.0033\,\mathrm{M}_{\sun}$ from \textit{AKARI} data \citep{MohMacLan12}, or possibly as small as $3.8\times10^{-4}\,\mathrm{M}_{\sun}$ \citep{CoxKervMarEA12}.
This is $\gtrsim20\times$ less massive than steady-state bow shocks ($M_s\sim 0.05-0.15\,\mathrm{M}_{\sun}$) from simulations of runaway RSGs \citep{MohMacLan12}, a problem which motivated the study presented here.
In our model the expanding RSG wind effectively encounters a vacuum initially, so there is no swept up shell from the early expansion and consequently its mass is much lower than a normal RSG bow shock,
$M_s \leq0.002\,\mathrm{M}_{\sun}$.
Our model therefore provides a good explanation for the mass of the bow shock, and is consistent with the RSG wind mass.

\textit{Herschel} observations of Betelgeuse \citep{CoxKervMarEA12} show that the bow shock seems to have a layered structure.
The local maximum in wind density during the BSG$\rightarrow$RSG transition (Fig.~\ref{fig:HRD}) remains visible in our simulation (Figs.~\ref{fig:DensTemp} and \ref{fig:DensityCuts}) for some time outside the inner shell, and such modulations in the wind properties could explain the observed layers.


There are two upstream discontinuities that could correspond to the bar seen upstream from Betelgeuse's bow shock, namely the BSG forward shock and contact discontinuity (or both, if the bow shock were a thin shell).
For the simulation presented here the contact discontinuity provides the most likely counterpart, based on its position and shape at the time the inner shocked-shell is visible.
We calculate the observed properties of the 2-D models (they have cylindrical symmetry) to test this hypothesis. The projected dust-emission intensity (upper half-plane, in $\mathrm{erg}\,\mathrm{s}^{-1}$ per pixel) and projected mass (lower half-plane, in $\mathrm{g}\,\mathrm{cm}^{-2}$) are shown in Fig.~\ref{fig:Projected} at time (C) (11.8875 Myr).
Dust emission in the IRAS $65\,\mu$m waveband is calculated assuming radiative equilibrium, and using the same assumptions for dust size and composition as \citet{NeiNgeKanEA10} with a dust-to-gas ratio of $1/200$.
The inner shocked-shell and a prominent outer shell are still apparent, both in projected density and in dust emission.
The $r^{-2}$ decrease in dust emission makes the contact discontinuity look a little more like a bar (compared to projected density) because the strongly curved parts at larger radii are fainter.
The total dust luminosity of the RSG wind, bow shock, and bar from this model are lower than the {\it{IRAS}}/{\it{AKARI}} observations, suggesting radiative excitation from Betelgeuse's radiation may not be the dominant process exciting the dust emission.
Converting from simulated mass density to far-infrared emission is complicated, however, because the dust composition of the ISM, BSG and RSG winds are probably all somewhat different.
Increasing the dust-to-gas ratio would increase the predicted luminosity.
It is also not clear what contribution C and O fine-structure lines make to the far-IR emission.
We have only considered Betelgeuse's radiation as the energy source for the dust emission, but radiation produced within the bow shocks and hot bubble probably also contribute.

The observed bar has no discernible curvature, but a strong prediction of our model is that it must eventually begin to curve back into a bow shock shape, if physically associated with Betelgeuse.
The radius of curvature can be larger than the current standoff distance of the bar would suggest if it is a relic structure that is currently being overtaken by the star (cf.\ Fig.~\ref{fig:DensTemp}). 
Instabilities provide another mechanism for reducing bow shock curvature (compare panels A and B of Fig.~\ref{fig:DensTemp}): the shock flexes periodically as vortices are generated and shed downstream, and the curvature of the contact discontinuity can consequently fluctuate quite dramatically.
This has also been seen in previous work \citep[e.g.][]{ComKap98,WarZijOBr07,VilManGar12}.

The transition timescale, from when the RSG wind appears to when the BSG wind's bow shock has disappeared, can be quite long and is beyond the end of the star's life in this model.
The BSG bow shock loses pressure support on a timescale comparable to the 10-15 kyr required for the star to expand to a RSG.
Subsequently, if the star is moving with $v_*=50\,\mathrm{km}\,\mathrm{s}^{-1}$ and the wind has $R_{\mathrm{SO}}=1\,$pc, the time to cross this distance is $\sim20\,$kyr but it can take three times as long for the bow shock to stall and be swept far downstream.
For larger $R_{\mathrm{SO}}$ and lower $v_*$ it is possible for the observable bow shock transition to last $\sim100$ kyr, a significant fraction of the massive star's post-MS evolution, hence it may not be unusual to observe a RSG during this transition.  
The transition from MS to RSG (not presented here) is qualitatively similar, differing in the wind strengths and speed of transition.

A number of uncertainties affect our model.  Betelgeuse's mass is poorly constrained \citep{NeiLesHau11} and the properties of the proposed blue phase of evolution are unknown, hence the BSG wind parameters are unconstrained.
Even with accurate stellar parameters, blue loop evolution is very sensitive to assumed physics in the models \citep[e.g.][]{NeiCanLan11}.
The ISM density and the ambient magnetic field (not considered here) also affect shock properties, and even Betelgeuse's spatial velocity through its local ISM has significant uncertainties \citep{UetIzuYamEA08}.
Stronger observational constraints on the ambient ISM properties would greatly help in constructing better models tailored specifically to Betelgeuse.
Multiwavelength observations of the bar and bow shock would better constrain their dust properties, providing more reliable mass estimates.
We expect that new IR observations from e.g.\ \textit{Herschel} and \textit{SOFIA} will strongly test and constrain our model.

\section{Conclusions} \label{sec:conclusions}
We have presented the first quantitative calculation of bow shock evolution for a massive runaway star undergoing the transition from blue to red on the HR diagram, here for a $15\,\mathrm{M}_{\sun}$ star evolving from BSG to RSG.
In contrast to the previously-studied case of the transition from RSG to WR, here the initial wind is faster than the following wind and so the bow shock has time to stall and collapse in on itself before it is destroyed and re-formed by the newly expanding RSG wind.

The BSG reverse shock collapses in on the RSG wind very rapidly (at the hot-bubble sound speed), producing an inner shocked-shell around the expanding RSG wind, and eventually wraps around to completely enclose it.
There is a short timeframe (10 kyr in this calculation) during which this inner shocked-shell resembles a bow shock and can be observed simultaneously with the much larger-scale bow shock left over from the BSG wind.
This bow shock is then overtaken by the slowly expanding RSG wind.
The transition time during which the BSG bow shock is still interacting with the expanding RSG wind can be $50-100$ kyr, depending on the
star's velocity through the ISM, the strength of the fast wind, and the ambient ISM density.
This is already a significant percentage of the time a massive star spends as a RSG, so it is possible that runaway RSGs may be commonly observed to have such features in their CSM.

The inner shocked-shell's mass, shape, thickness and size are consistent with the observed properties of Betelgeuse's bow shock, in contrast to steady-state bow shocks which are typically more than an order of magnitude more massive.
Our model also naturally provides a possible counterpart to the infrared bar ahead of Betelgeuse's bow shock, the BSG wind's shocked-shell.
We have also shown that dust emission from the inner shocked-shell and outer bow shock may have similar emissivities.

\begin{acknowledgements}
The authors are grateful to Vasilii Gvaramadze for providing Fig.~\ref{fig:Betelgeuse}.
JM and HN are funded by fellowships from the Alexander von Humboldt Foundation.
This work was supported by the Deutsche Forschungsgemeinschaft priority program 1573, ``Physics of the Interstellar Medium''.
The authors acknowledge the John von Neumann Institute for Computing for a grant of computing time on the JUROPA supercomputer at J\"ulich Supercomputing Centre.
\end{acknowledgements}

\bibliographystyle{apj}



\end{document}